\documentclass{moriond}

\def\GeV{\ensuremath{~\textrm{GeV}}}
\def\TeV{\ensuremath{~\textrm{TeV}}}
\def\fbinv{\ensuremath{~\textrm{fb}^{-1}}}
\def\pt{\ensuremath{{p}_{T}}}
\def\PZpr{\ensuremath{\textrm{Z}^\prime}}
\def\PWpr{\ensuremath{\textrm{W}^\prime}}
\def\bbbar{\ensuremath{\textrm{b}\bar{\textrm{b}}}}
\def\qqbar{\ensuremath{\textrm{q}\bar{\textrm{q}}}}

\def\be{\begin{equation}}
\def\ee{\end{equation}}
\def\bea{\begin{eqnarray}}
\def\eea{\end{eqnarray}}

\newcommand{\Photo}{\includegraphics[height=35mm]{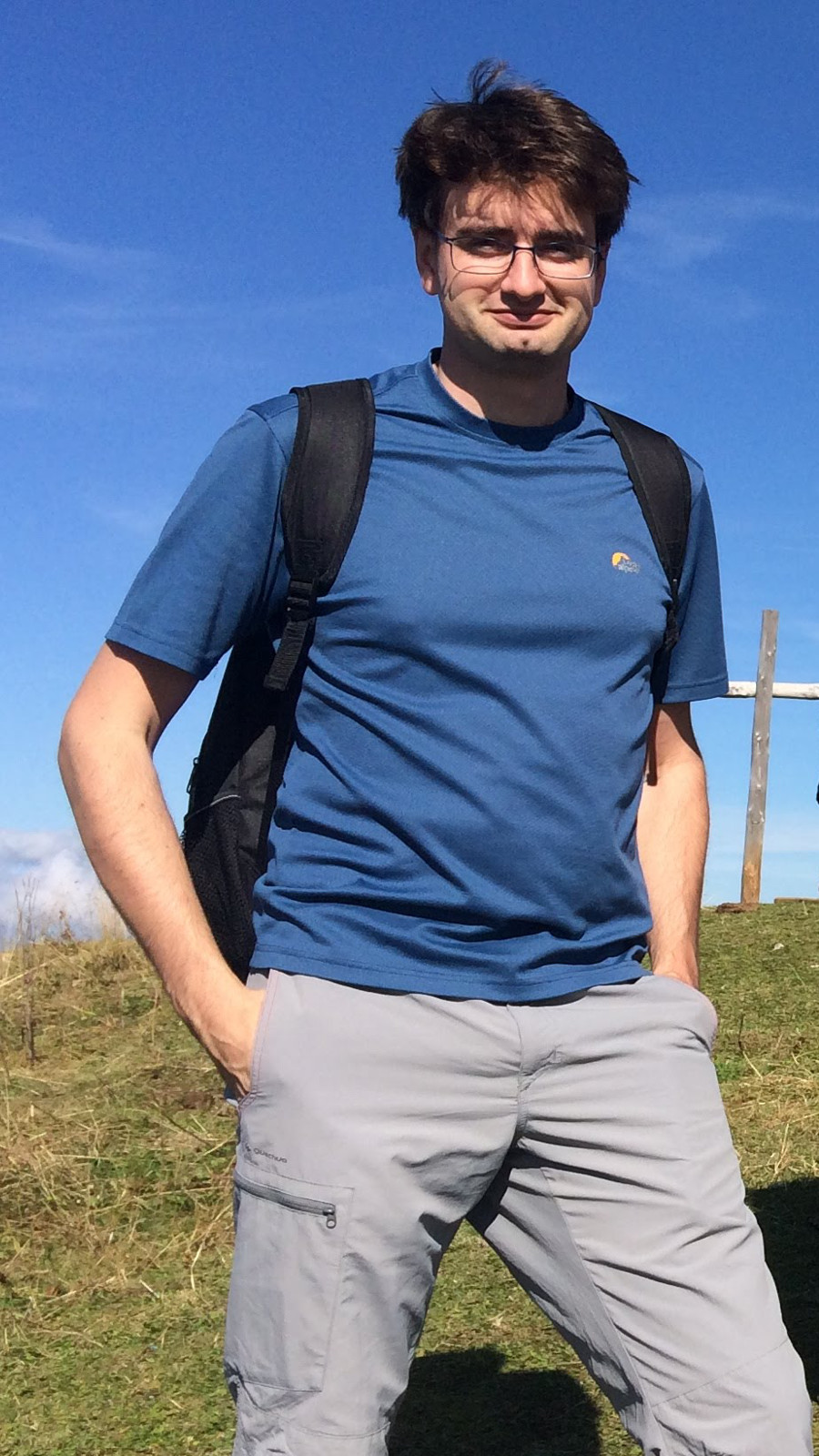}}

\begin{document}
\vspace*{4cm}
\title{Heavy resonances (\PWpr, \PZpr, jets) in ATLAS and CMS in Run 2}

\author{A. Zucchetta on behalf of the ATLAS and CMS Collaborations}

\address{Universit\"at Z\"urich, Physik-Institut, Winterthurerstrasse 190, 8057 Z\"urich}

\maketitle\abstracts{
An overview of the searches for heavy resonances that decay to leptons or quarks is presented. The results are based on the data collected by the ATLAS and CMS experiments in proton-proton collisions at $\sqrt{s}=13 \TeV$, and include for the first time the entire LHC Run 2 data set.
}

\footnotetext{\textsf{\copyright~2019 CERN for the benefit of the ATLAS and CMS Collaborations. CC-BY-4.0 license.}}

\section{Introduction}

New resonances are predicted by several theories addressing the open issues of the standard model (SM). Depending on the mass and the couplings to the SM quarks and leptons of the heavy particles, these new states could be accessible to the LHC and observable by the ATLAS~\cite{atlas} and CMS~\cite{cms} experiments. Spin-1 heavy resonances are predicted in minimal extensions of the SM, as in the sequential standard model (SSM) $\PWpr$~\cite{Wprime} and $\PZpr$~\cite{Zprime}. 
Massive spin-2 gravitons (G) may represent the first Kaluza-Klein excitation in Randall-Sundrum warped extra dimension models~\cite{rs}. Exotic models also include vector-like quarks (VLQ)~\cite{vlq}, and excited leptons in compositeness models~\cite{compo}.

As the most powerful proton-proton collider currently in operation, the LHC has the possibility to make such a discovery. At the end of 2018, the LHC Run 2 has finished its operations, and thus a large data set, accounting for approximately $140\fbinv$, is now available for analysis. In this proceeding, the first results based on the full Run 2 data set are presented.

\section{Dilepton searches}\label{subsec:dilepton}

The dilepton (ee or $\mu\mu$) final state is considered the golden channel for the discovery of a heavy resonance, thanks to the minimal backgrounds from the SM, and the electron and muon reconstruction efficiency and energy resolution. The ATLAS experiment presents the first search for dilepton resonances based on the full Run 2 data set~\cite{atlas_dilepton}, which amounts to 139\fbinv. The CMS experiment has updated the dielectron search to $77\fbinv$ collected during 2016 and 2017~\cite{cms_dielectron}, while the dimuon channel is based on the 2016 data set consisting of $36\fbinv$ of data~\cite{cms_dimuon}.

These searches select events with two same-flavor leptons, with a large transverse momentum (\pt), and isolated from other hadronic activity in the event. In ATLAS, the electron channel dominates the sensitivity, because of the larger selection efficiency with respect to muons (70--75\% and 50--40\%, respectively) and the far better dilepton resolution (a flat $\approx 1\%$ for electrons, and 3--27\% for muons). In CMS, the dielectron reconstruction efficiency is 70\%, and the dielectron invariant mass resolution is between 1.3 and 2.0\%. The dimuon selection efficiency is larger at 93\%, and the momentum resolution is about 7\% for muons with \pt~of 1\TeV.

\begin{figure}[!htb]
\begin{minipage}{0.575\linewidth}
\centerline{\includegraphics[width=\linewidth]{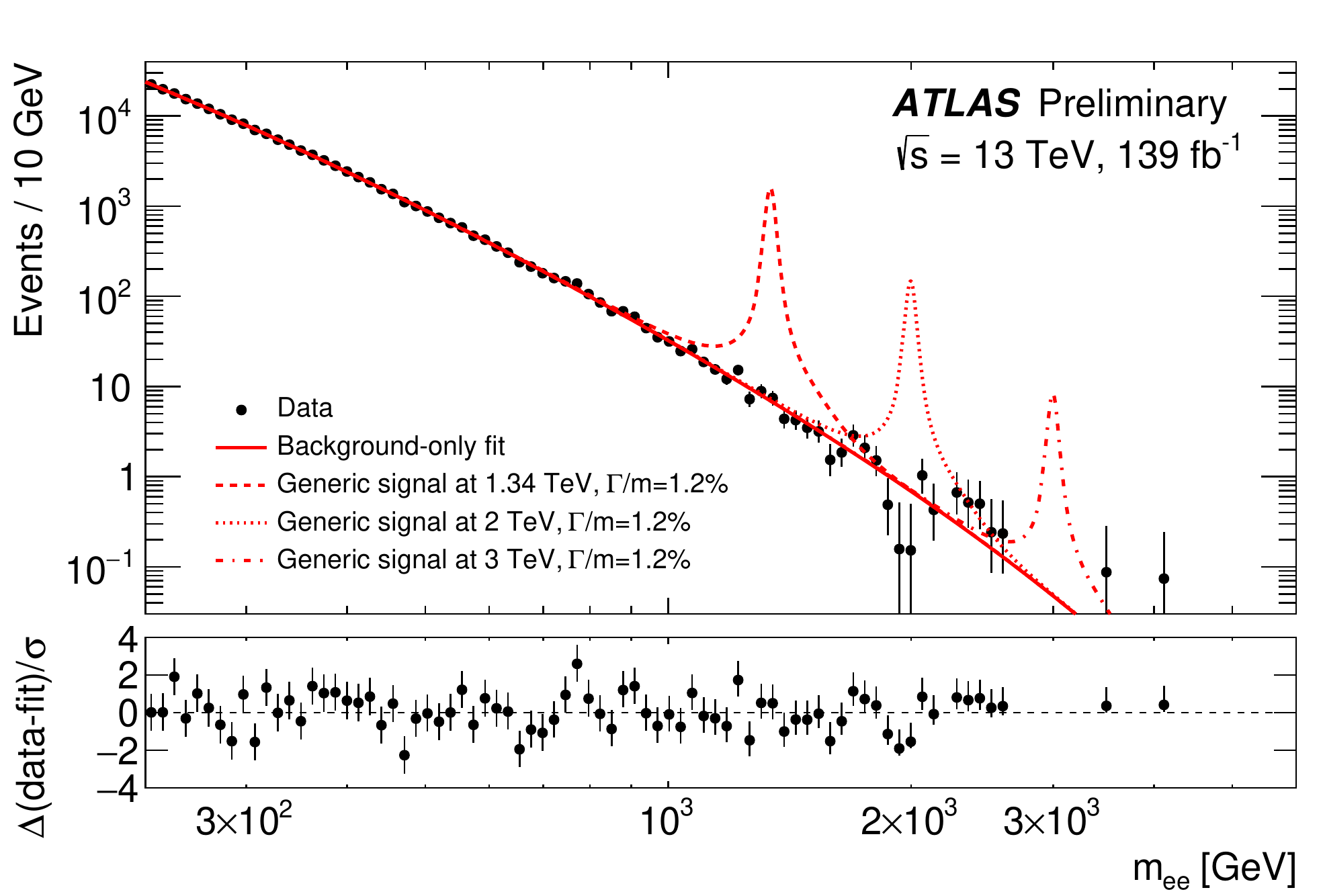}}
\end{minipage}
\hfill
\begin{minipage}{0.415\linewidth}
\centerline{\includegraphics[trim=0 0 240 0,clip,width=\linewidth]{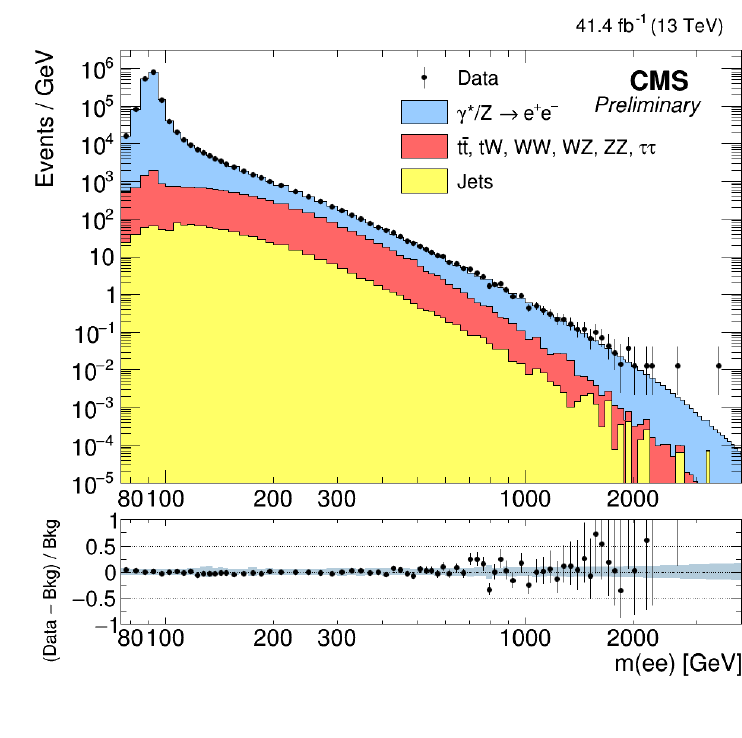}}
\end{minipage}
\hfill
\caption[]{Dielectron invariant mass observed by ATLAS~\cite{atlas_dilepton} with $139\fbinv$ (left) and by CMS~\cite{cms_dielectron} with $77\fbinv$ (right).}
\label{fig:dielectron}
\end{figure}

A heavy resonance would emerge as a localized excess of events above the smoothly falling SM background processes.
The SM background, represented mostly by off-shell Z bosons, is estimated in ATLAS by performing a combined fit of the signal model and a smooth power-law function that is used to parametrize the background distribution. The signal model is composed by a variable-width non-relativistic Breit-Wigner function modeling the natural width of the resonance, and a sum of a Gaussian and a Crystal Ball function to describe the detector response. The background estimation of the CMS search is instead based on Monte Carlo simulations of the background processes. The dielectron invariant mass spectra are presented in Fig.~\ref{fig:dielectron}.

\begin{figure}[!htbp]
\begin{minipage}{0.495\linewidth}
\centerline{\includegraphics[width=\linewidth]{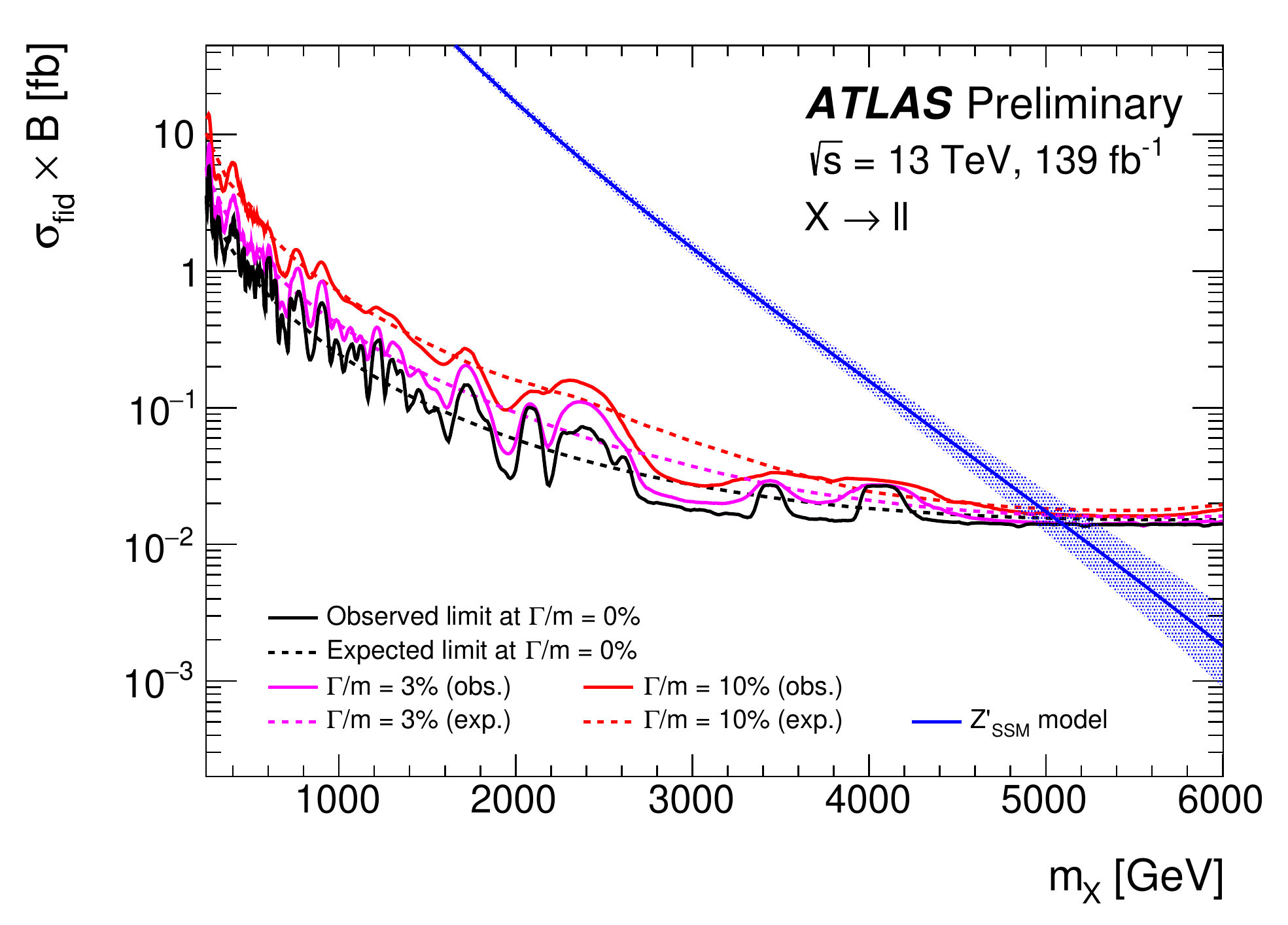}}
\end{minipage}
\hfill
\begin{minipage}{0.495\linewidth}
\centerline{\includegraphics[width=\linewidth]{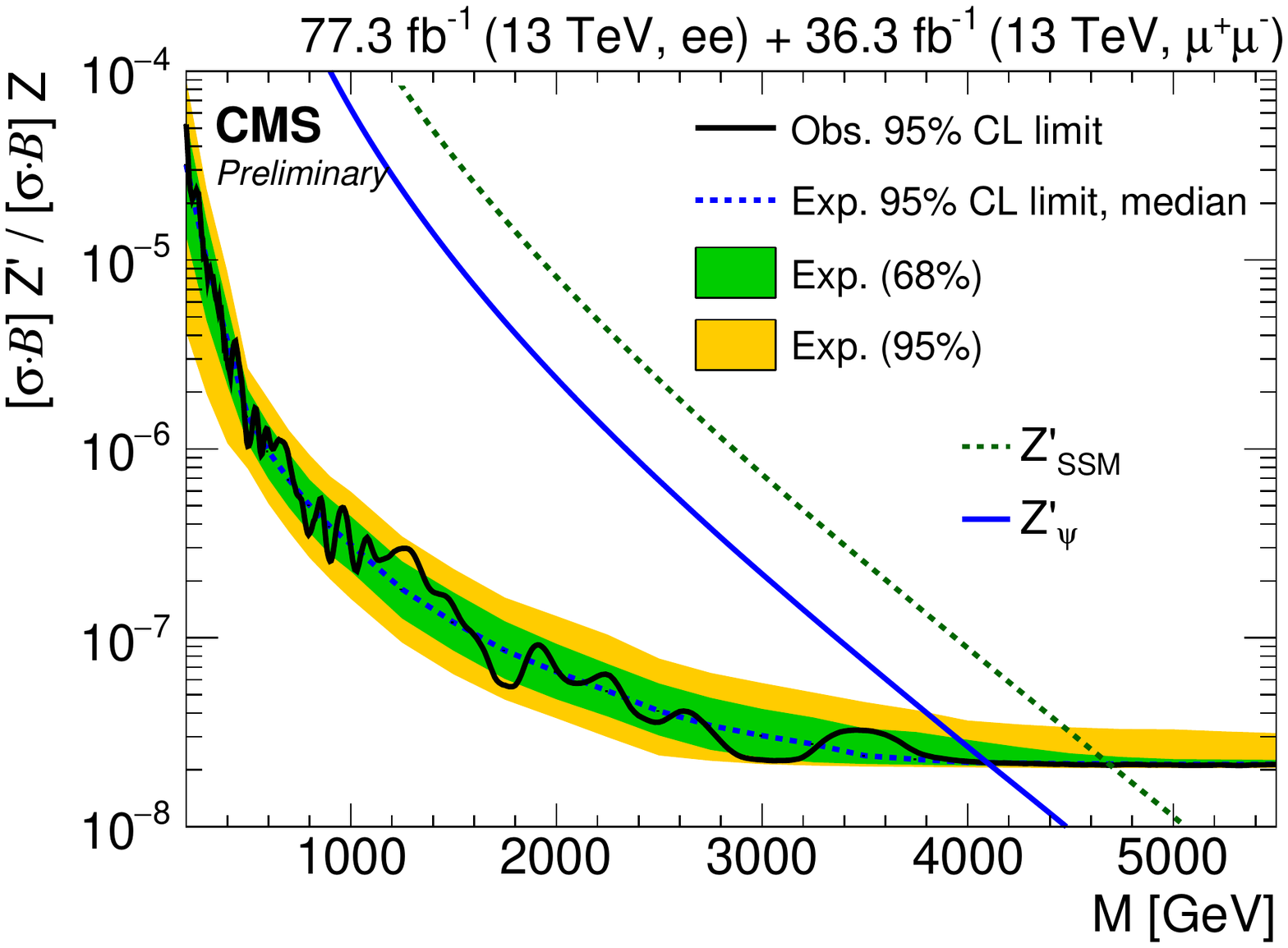}}
\end{minipage}
\hfill
\caption[]{Exclusion limits from ATLAS~\cite{atlas_dilepton} (left) and CMS~\cite{cms_dielectron} (right) on heavy resonances decaying to dilepton final states. The result represent the combination of the ee and $\mu\mu$ channels.}
\label{fig:dilepton}
\end{figure}

Both experiments report no significant excesses above the SM background expectation. The most significant local excess observed by ATLAS corresponding to the dielectron invariant mass of 774\GeV~amounts to 2.9 standard deviations, which become 0.1 when the \emph{look-elsewhere} effect is taken into account. Upper limits at 95\% confidence level (CL) are set on the cross section of heavy resonances decaying to a pair of leptons, shown in Fig.~\ref{fig:dilepton}. The ATLAS search also performs a scan of the resonance width up to 10\% of its mass. The results reported by ATLAS exclude a SSM $\PZpr$ with mass lower than 5.1\TeV, and CMS, in spite of the smaller data set, is able to exclude up to 4.7\TeV.

\section{Dilepton combination}\label{subsec:combo}

\begin{figure}[!htb]
\begin{minipage}{0.495\linewidth}
\centerline{\includegraphics[width=\linewidth]{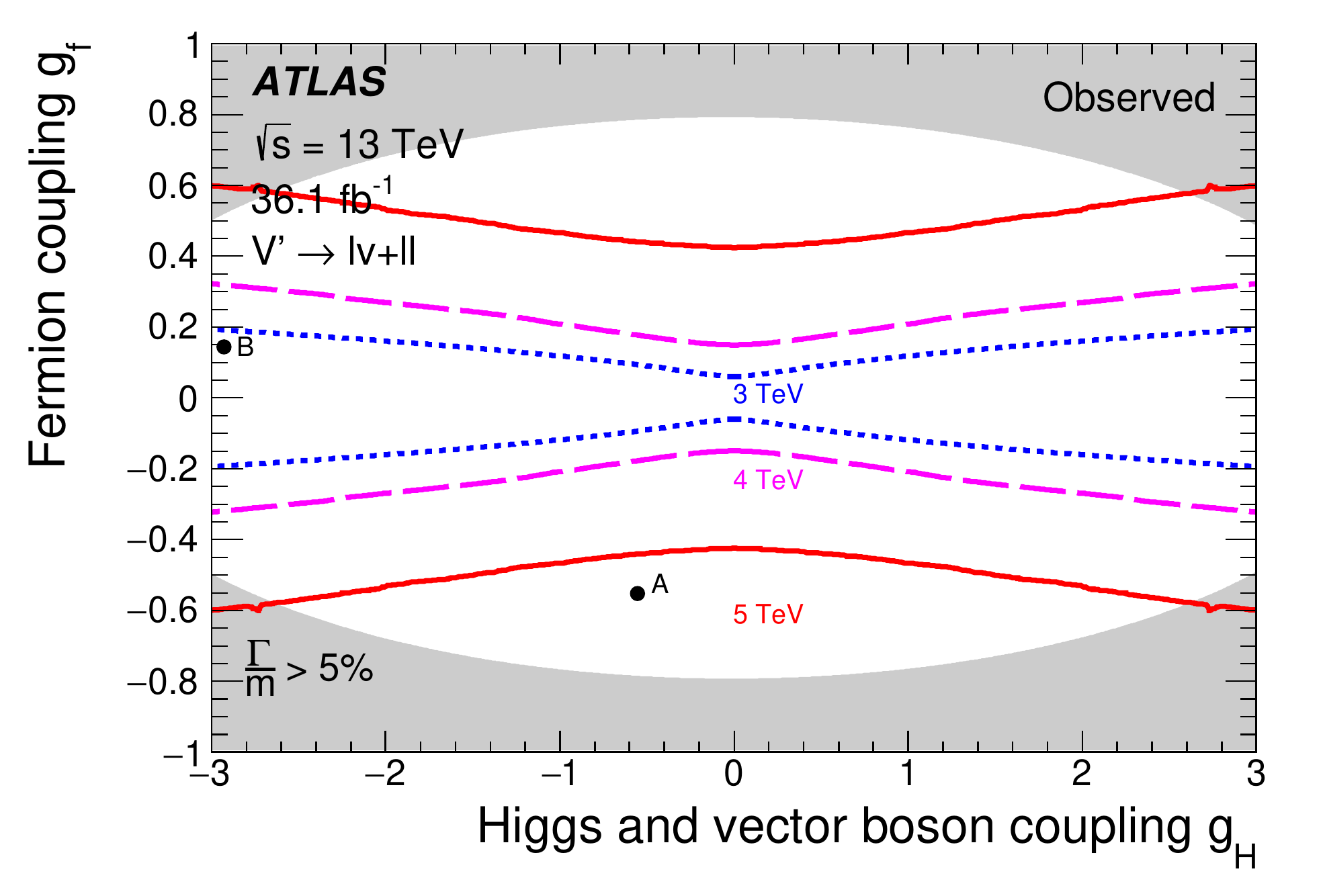}}
\end{minipage}
\hfill
\begin{minipage}{0.495\linewidth}
\centerline{\includegraphics[width=\linewidth]{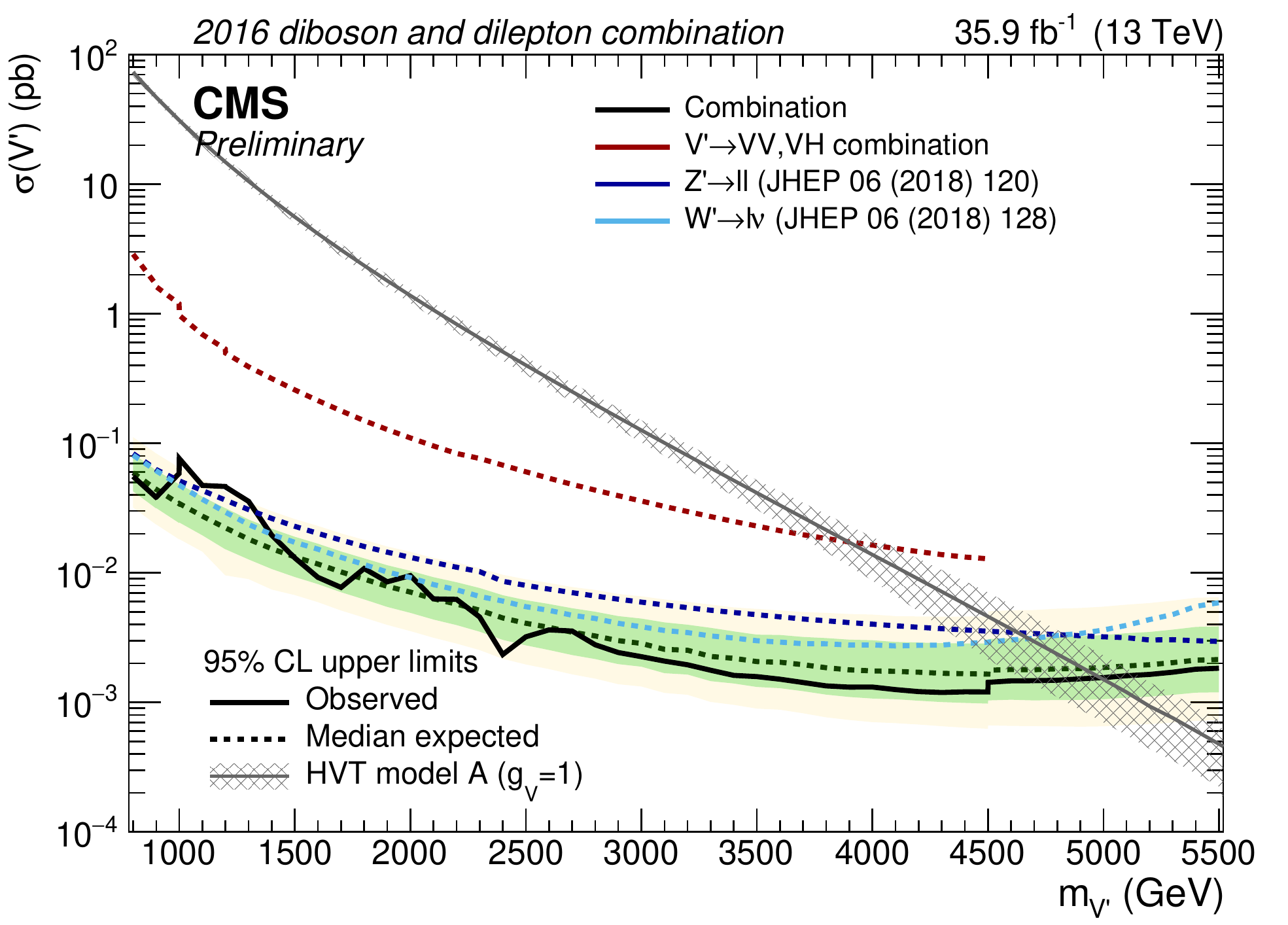}}
\end{minipage}
\hfill
\caption[]{Exclusion limits from ATLAS in the plane of the couplings to fermions and SM bosons of a triplet of heavy resonances~\cite{atlas_combo} (left). Exclusion limit from CMS by combining the $\PWpr\to \ell\nu$ and $\PZpr\to\ell\ell$ searches in a scenario where the couplings of the heavy resonances are SM-like~\cite{cms_combo} (right).}
\label{fig:combo}
\end{figure}

Even though the minimal extensions of the SM predict only one heavy resonance, other theories, described in an effective approach by the heavy vector triplet (HVT) model~\cite{Pappadopulo2014}, foresee the simultaneous presence of mass-degenerate triplet of heavy resonances: one $\PZpr$ and two electrically charged $\PWpr$$^\pm$. In the scenarios with SM-like couplings, the $\PWpr\to \ell\nu$ and $\PZpr\to\ell\ell$ decays represent the privileged channels for discovery. The ATLAS and CMS Collaborations recently published a statistical combination of these final states based on the 2016 data set~\cite{atlas_combo}~\cite{cms_combo}, which is complemented by the $\PWpr$ and $\PZpr$ decays to a pair of SM bosons in the scenarios where the couplings to bosons are enhanced with respect to those to fermions. The dilepton and diboson searches provide complementary sensitivity, as dilepton searches constrain the fermionic couplings in regions that are not accessible to diboson final states alone, as depicted in Fig.~\ref{fig:combo}.

\section{Dijet searches}\label{subsec:dijet}

\begin{figure}[!htb]
\begin{minipage}{0.32\linewidth}
\centerline{\includegraphics[width=\linewidth]{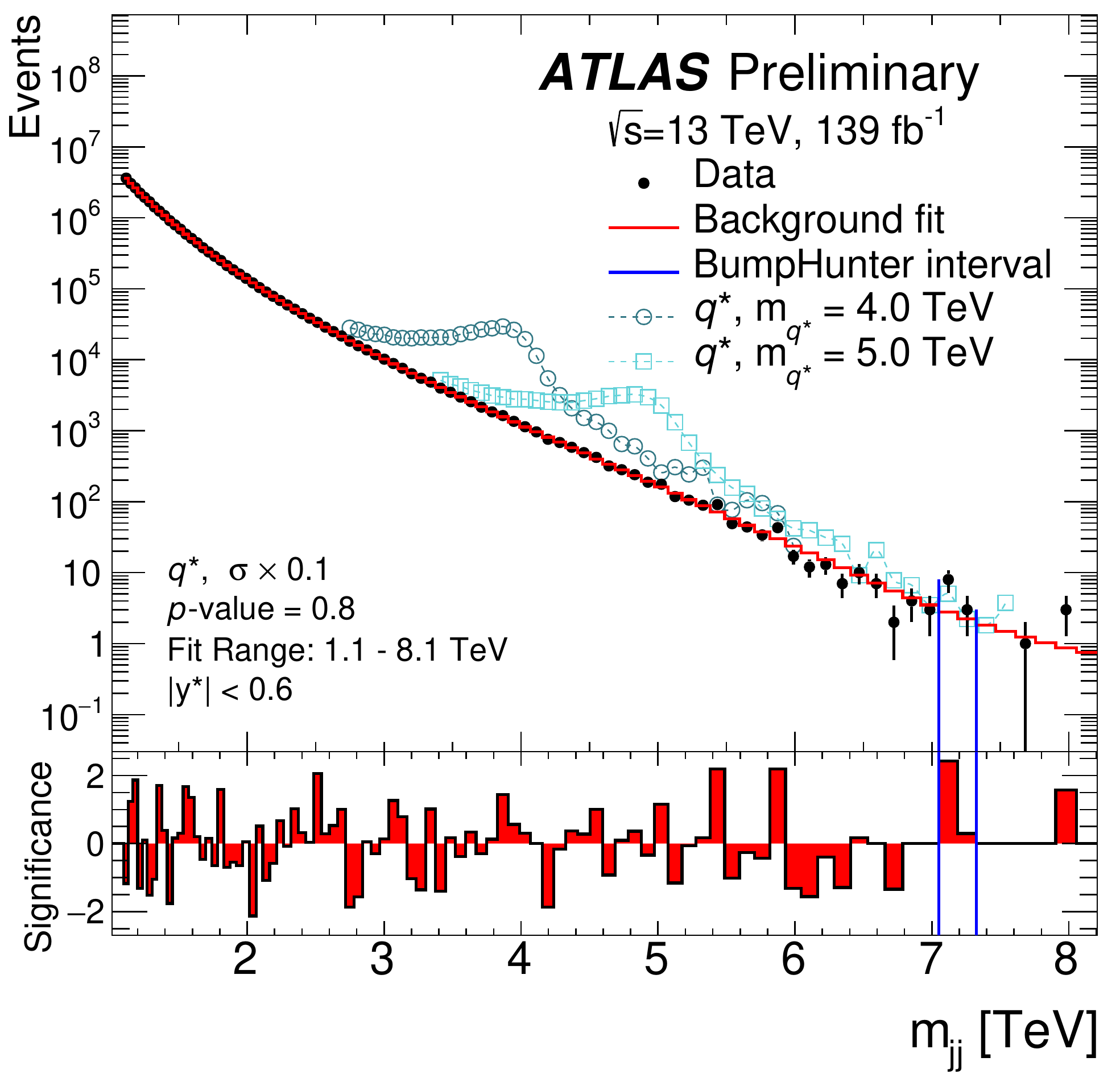}}
\end{minipage}
\begin{minipage}{0.30\linewidth}
\centerline{\includegraphics[width=\linewidth]{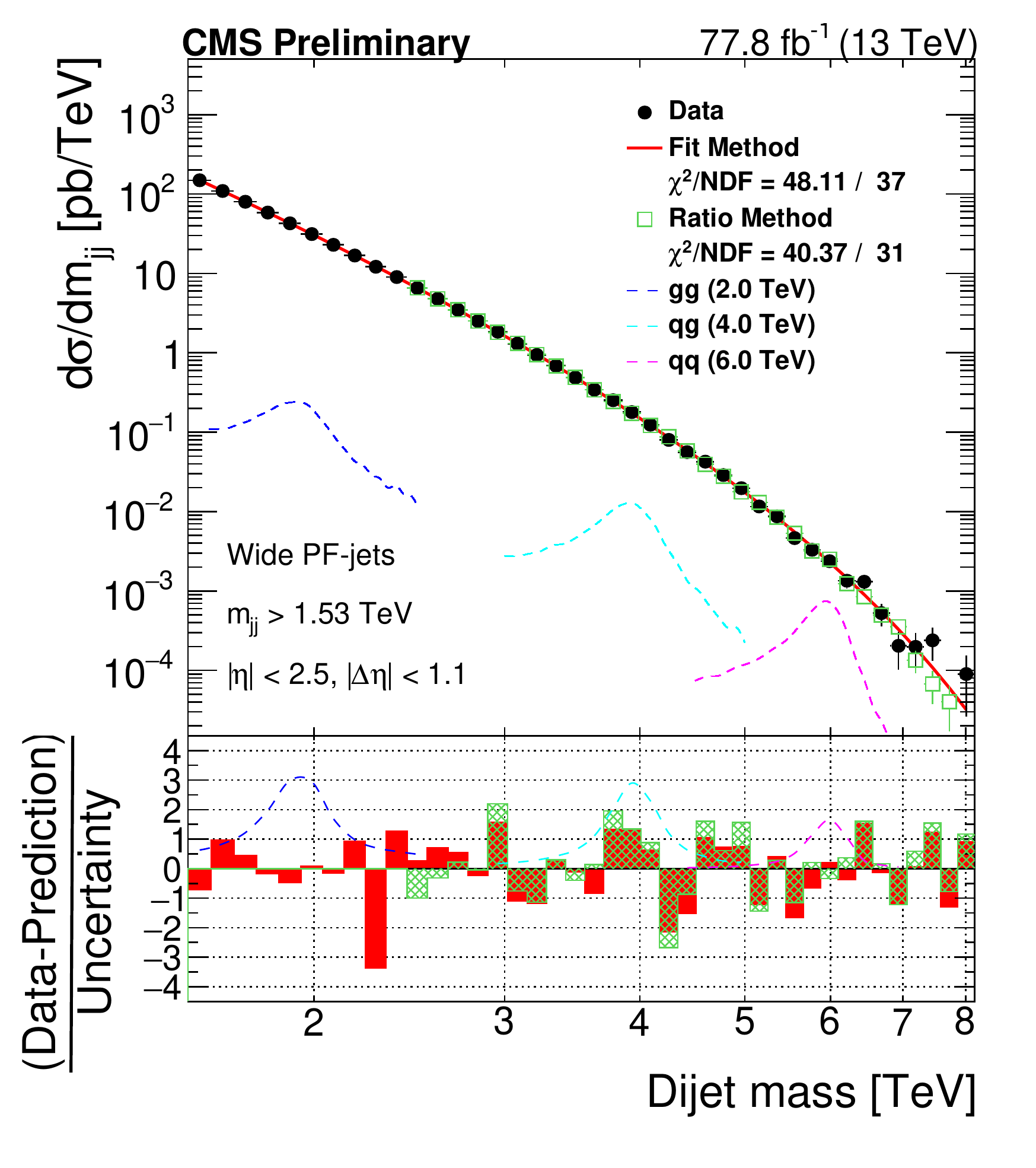}}
\end{minipage}
\hfill
\begin{minipage}{0.36\linewidth}
\centerline{\includegraphics[width=\linewidth]{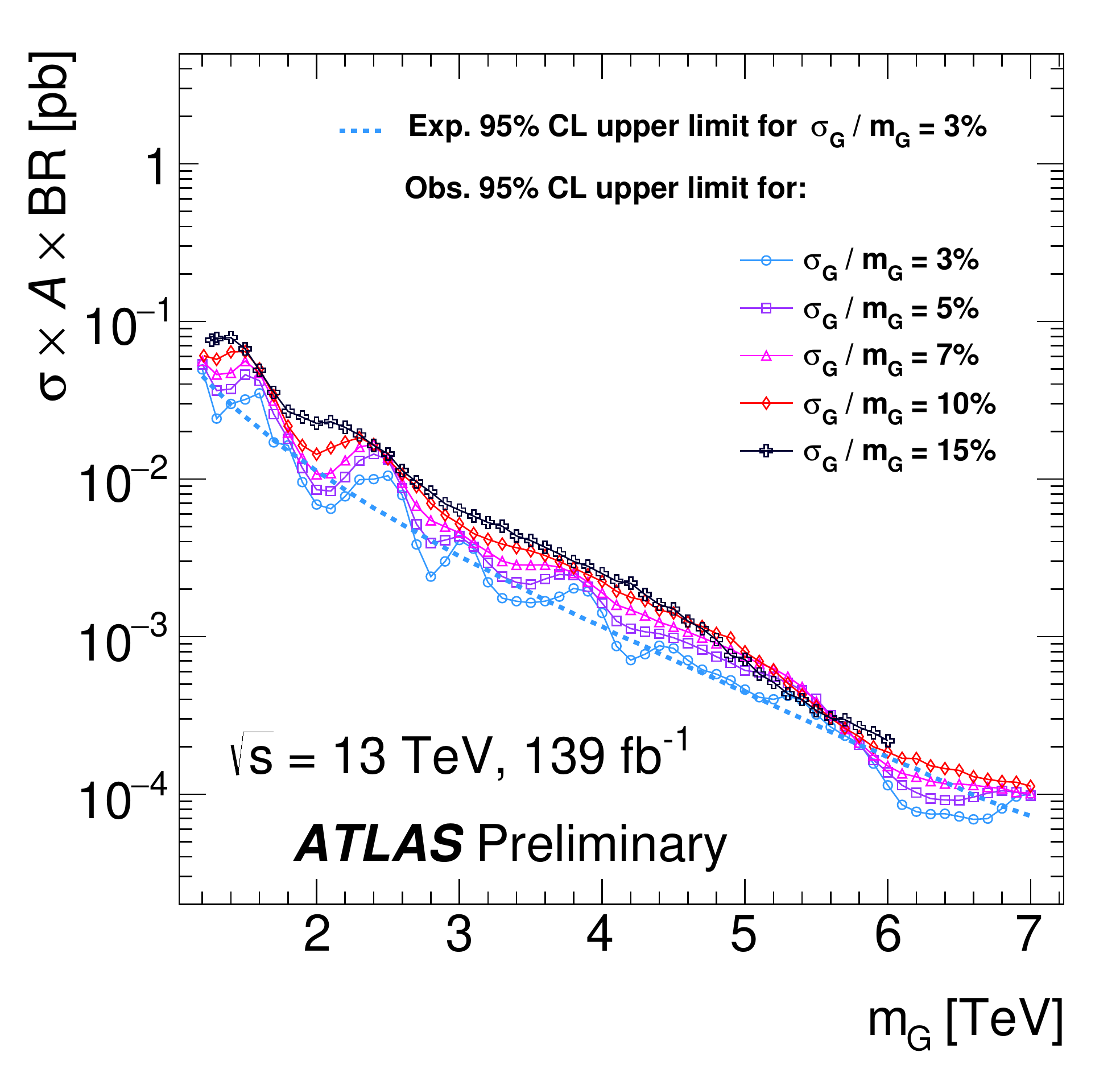}}
\end{minipage}
\hfill
\caption[]{Dijet invariant mass spectrum observed by ATLAS with the full Run 2 data set~\cite{atlas_dijet} (left); comparison between the CMS dijet invariant mass distribution in data and the two background models~\cite{cms_dijet} (center); ATLAS exclusion limits on a graviton with a natural width comprised between 3 and 15\% (right).}
\label{fig:dijet}
\end{figure}

Heavy resonances decaying to a pair of quarks or gluons represent one of the most natural searches for the LHC programme, as the initial and final states are the same. The ATLAS experiment recently published the results for dijet resonances based on the full Run 2 data set~\cite{atlas_dijet}, accounting for 139\fbinv. This search is favored by the dijet invariant mass resolution, which is stable between 2.4--2.9\% in the 1.2--8.1\TeV~range probed in this analysis. On the other hand, the QCD multijet production background is very large; its $t$-channel dijet production mode is suppressed by requiring a minimal separation in the rapidity angle between the two jets, $\frac{1}{2} | y_{j1} - y_{j2} | < 0.6$. The SM background is estimated by parametrizing its distribution with a power-law function with 4 free parameters, as reported in Fig.~\ref{fig:dijet}. The results are expressed as upper limits on the product of the cross section, acceptance, and selection efficiency as a function of the resonance mass and width, the latter varying from 3 to 15\%. In the context of excited quarks, the exclusion limit is increased by 700\GeV~with respect to the previous publications.


The corresponding CMS search is based on $77\fbinv$ of data~\cite{cms_dijet}. In the region with dijet invariant mass larger than 2.4\TeV, the analysis introduces a novel background estimation technique based on three regions defined depending on the separation $\Delta \eta$ between the two jets. The background in the signal region ($\Delta \eta < 1.1$) is estimated from a control region ($\Delta \eta > 1.5$) and a transfer function based on simulation. The intermediate validation region is used to derive a further correction from the data in the signal region. This technique yields results consistent with the previous function fit method but is more robust against non-narrow resonances that are reconstructed as broad excesses of events. Additionally, in order to increase the dijet resolution and recover particles emitted from final state radiation, jets are merged to the two leading jets if the angular separation between them is small enough ($\Delta R = \sqrt{\Delta \eta^2 + \Delta \phi^2} < 1.1$).
No significant excess is seen, as reported in Fig.~\ref{fig:dijet}, and upper limits are set on the cross section of the resonances predicted by several theories beyond the SM. 


\section{Low mass dijet searches}\label{subsec:lowmass}

The sensitivity of high-mass dijet searches is inferiorly limited by the trigger thresholds, which prevent the exploration of the sub-TeV regime. This limitation can be overcome by requiring more complex final states involving additional particles, such as an initial state radiation (ISR) jet or a photon, or alternatively, if the resonance decays to b quarks, by tagging its heavy flavor content. The changes in the topology imply that the resonance recoils against the ISR particles and is produced with a large boost.

The ATLAS Collaboration presents a search based on 80\fbinv~for light resonances decaying to a pair of quarks or b quarks in association with a photon~\cite{atlas_gbb}. Two triggers based on high energy photons are used to collect data, and two categories (flavor inclusive and b tagged) enhance the sensitivity to $\PZpr\to\bbbar$ decays. These selections lower the probed \PZpr~masses down to 225\GeV.

Both ATLAS and CMS are looking for light resonances decaying to b quarks in region of the phase space with large resonance boost~\cite{atlas_jbb}~\cite{cms_jbb}. Both experiments use large-cone jets (with radius parameter 1.0 for ATLAS, and CMS uses either 0.8 or 1.5) and the investigation of the jet substructure to simultaneously identify the two b quarks within the same jet. The events are required to have at least one additional high-\pt~jet, and the exotic resonance (a \PZpr~or a scalar $\Phi$) is sought in the large-cone jet mass spectrum. The usage of jet substructure extends the resonances masses down to approximately the W/Z boson masses.

A significant improvement in this field is presented by CMS exploring low mass resonances decaying to a pair of light quarks~\cite{cms_gjj}. The events are collected by triggering on an ISR photon, and the dominating non-resonant background from $\gamma$+jets is estimated using a novel technique based on the Energy Correlation Functions~\cite{d2} decorrelated from the jet mass. Upper limits on the cross section of light $\PZpr$ resonance or on its couplings to quarks can be set for masses as low as 10\GeV. The corresponding ATLAS search~\cite{atlas_gjj} uses a N-subjettiness $\tau_{21}$ tagger~\cite{tau21}, also decorrelated from the jet mass, but stops at 100\GeV.

\begin{figure}[!htb]
\begin{minipage}{0.38\linewidth}
\centerline{\includegraphics[width=\linewidth]{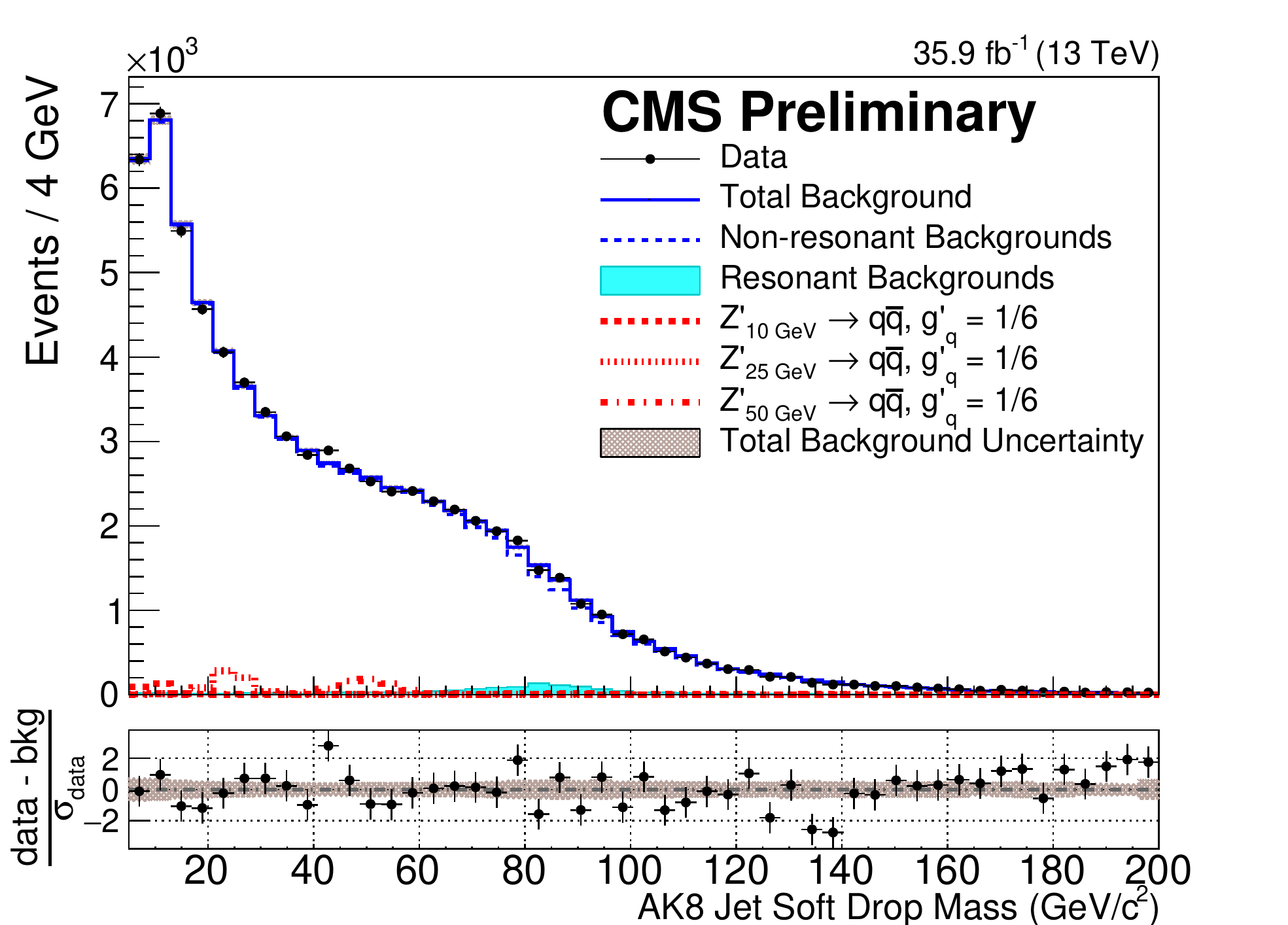}}
\end{minipage}
\begin{minipage}{0.6\linewidth}
\centerline{\includegraphics[width=\linewidth]{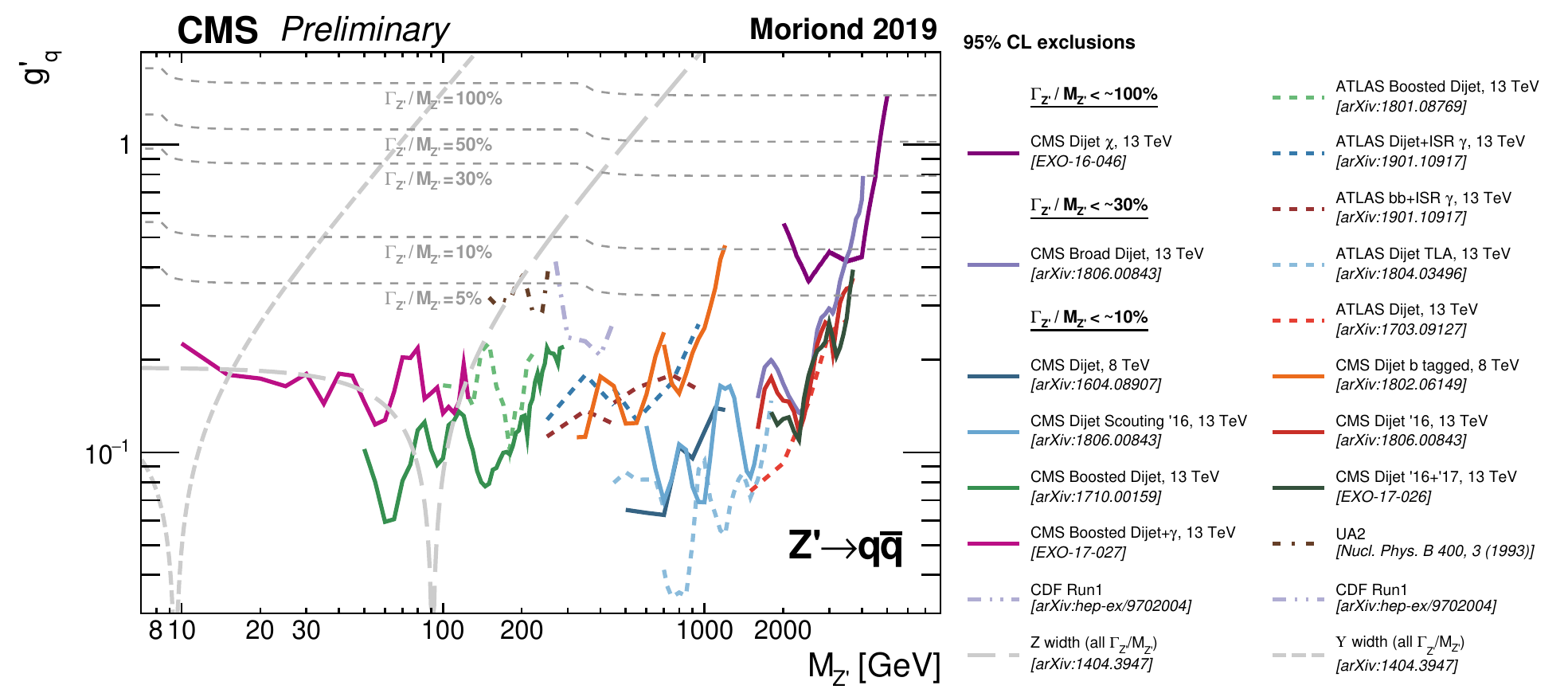}}
\end{minipage}
\hfill
\caption[]{Left: large-cone jet mass spectrum for events passing the analysis selections of the CMS search~\cite{cms_gjj}. Right: summary of the ATLAS and CMS dijet searches~\cite{dijet}, from 10\GeV~to 8\TeV.}
\label{fig:lowmass}
\end{figure}

These searches allow to extend the reach of the resonance searches at LHC to masses that would be naturally inaccessible by ATLAS and CMS, and cover the range from 10\GeV~to 8\TeV, as reported in Fig.~\ref{fig:lowmass}. However, no convincing excess is observed.

\section{Vector-like Quarks}\label{subsec:vlq}

Vector-like quarks (VLQ) are hypothetical heavy fermions that decay to a third-generation quark (t, b) and a SM boson (W, Z, H). The large mass of the VLQ implies that these particle have  large momenta, and can be reconstructed as single large-cone jets if they decay to quarks. Since VLQ are produced in pairs at the LHC, there is a large multiplicity of channels with a large number of possible final states. The CMS search~\cite{cms_vlq} is based on a novel tagger (Boosted Event Shape Tagger, or BEST) based on a deep neural network that is able to classify W, Z, H bosons, top and bottom quarks that decay to hadrons. Depending on the possible final states, 126 search regions are defined according to the BEST output. In each region, a fit to the scalar sum of the \pt~of the jets in the event is performed. The search, based on 36\fbinv~of data, extends the excluded mass for VLQ by approximately 100\GeV~with respect to the previous analyses, depending on the channel. The ATLAS experiment published a statistical combination of 6 different searches looking for VLQ in both fully-hadronic, semi-leptonic, and fully-leptonic final states~\cite{atlas_vlq}, as depicted in Fig.~\ref{fig:vlq}. The SM backgrounds are found to be consistent with the background expectation, and no significant excess is found.

\section{Excited leptons}\label{subsec:ex}

\begin{figure}[!htb]
\begin{minipage}{0.43\linewidth}
\centerline{\includegraphics[width=0.8\linewidth]{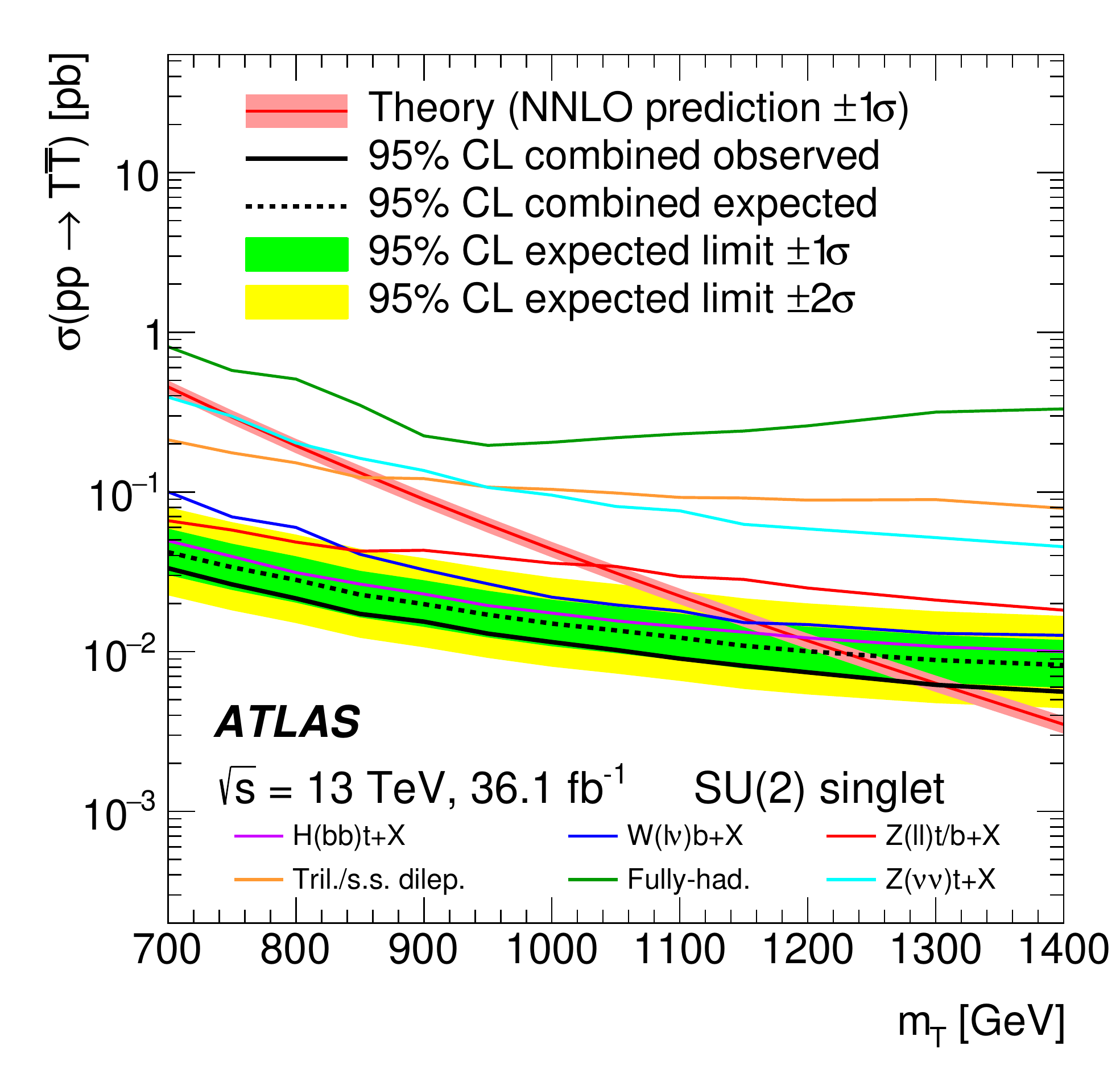}}
\end{minipage}
\begin{minipage}{0.55\linewidth}
\centerline{\includegraphics[width=0.8\linewidth]{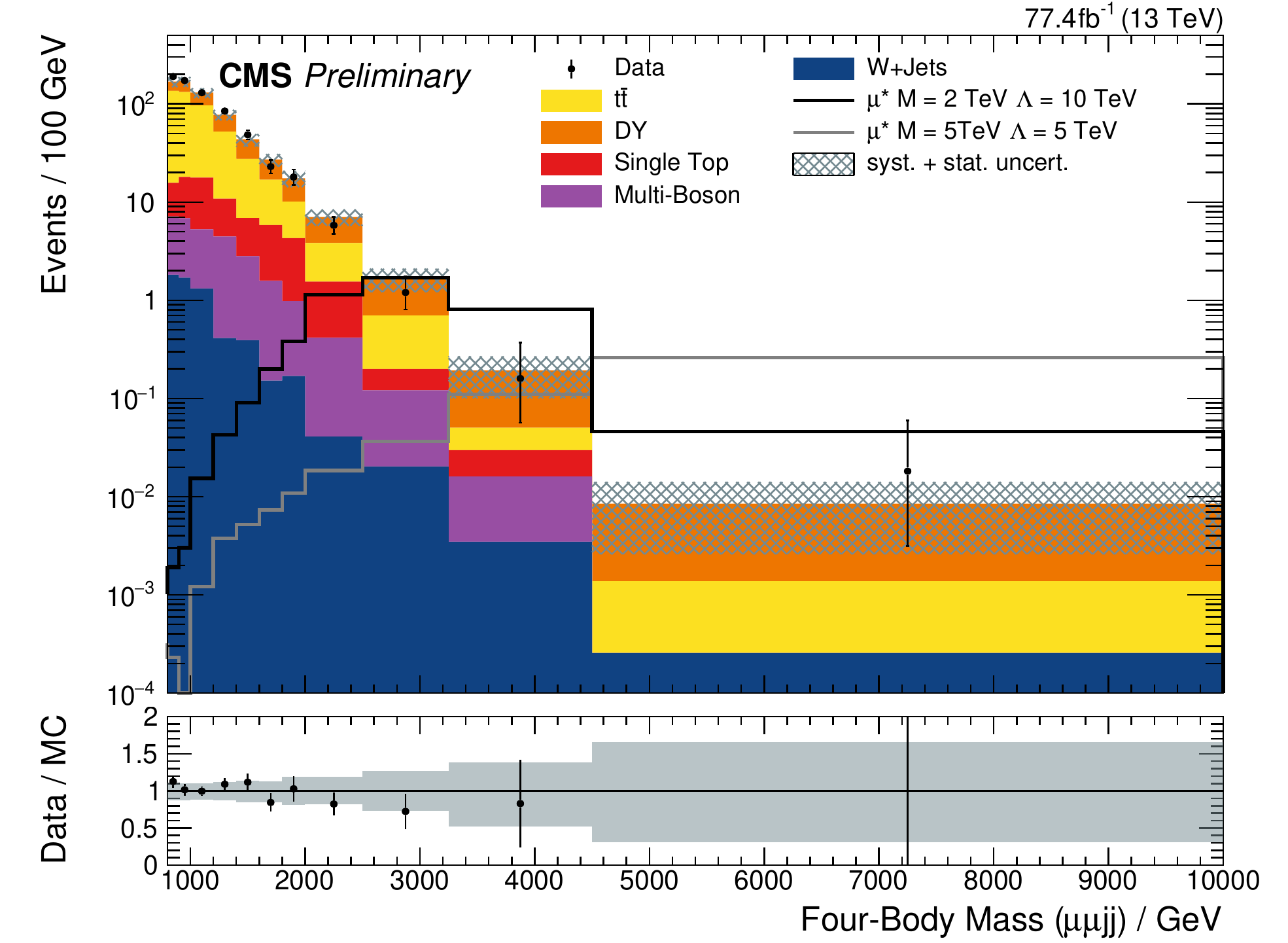}}
\end{minipage}
\hfill
\caption[]{Left: ATLAS combination of VLQ searches~\cite{atlas_vlq}. Right: four-body invariant mass spectrum of the CMS excited muon search~\cite{cms_lstar}.}
\label{fig:vlq}
\end{figure}

Excited leptons ($\ell^*$) are included in compositeness models, which postulate that leptons are not fundamental particles. Heavy excited leptons are thus produced from \qqbar~annihilation through contact interactions with scale $\Lambda$ and in association with a SM lepton. In the CMS search, which analyzes 77\fbinv~of data, the excited lepton also decays through contact interaction to an other SM lepton and a pair of quarks~\cite{cms_lstar}. The final state thus consists of two same-flavor leptons and to resolved jets. The main Z+jets background is estimated in the region where the Z boson is on-shell, while the search regions considers a dilepton invariant mass larger than 500\GeV. A fit is then performed on the four-body invariant mass, shown in Fig.~\ref{fig:vlq}. Assuming $\Lambda=M_{\ell^*}$, excited electrons and muon with mass smaller than 5.6 and 5.7\TeV~are excluded, respectively.

\section{Summary}\label{subsec:sum}

The data set provided by the LHC in the past four years of running is now being analyzed, and the first analyses looking for heavy dijet and dilepton resonances and exploiting the full integrated luminosity have been presented at the Moriond EW conference. However, including new data is not the only improvement from the ATLAS and CMS experiments. Searches are being refined by including more complex and robust background estimation methods, and designing bleeding edge taggers based on the latest machine learning developments. More searches will follow in the near future based on the new data and analysis techniques, significantly improving the reach of the LHC in the search for heavy resonances.

\end{document}